\documentclass[sigconf, authorversion,nonacm]{acmart}
\hypersetup{
	colorlinks=true,
	linkcolor=blue,
	filecolor=magenta,      
	urlcolor=cyan,
}

\AtBeginDocument{%
  \providecommand\BibTeX{{%
    \normalfont B\kern-0.5em{\scshape i\kern-0.25em b}\kern-0.8em\TeX}}}

\begin{document}

\title{A Transformer Architecture for Stress Detection from ECG}

\author{Behnam Behinaein}
\affiliation{%
	\institution{Department of Electrical and Computer Engineering, \\ Ingenuity Labs Research Institute}
	\streetaddress{}
	\city{Queen's University, Kingston}
	\state{}
	\country{Canada}
}
\email{9hbb@queensu.ca}

\author{Anubhav Bhatti}
\affiliation{%
	\institution{Department of Electrical and Computer Engineering, \\ Ingenuity Labs Research Institute}
	\streetaddress{}
	\city{Queen's University, Kingston}
	\state{}
	\country{Canada}
}
\email{anubhav.bhatti@queensu.ca}

\author{Dirk Rodenburg}
\affiliation{%
	\institution{Ingenuity Labs Research Institute}
	\streetaddress{}
	\city{Queen's University, Kingston}
	\state{}
	\country{Canada}
}
\email{djr08@queensu.ca}

\author{Paul Hungler}
\affiliation{%
	\institution{Ingenuity Labs Research Institute}
	\streetaddress{}
	\city{Queen's University, Kingston}
	\state{}
	\country{Canada}
}
\email{paul.hungler@queensu.ca}

\author{Ali Etemad}
\affiliation{%
	\institution{Department of Electrical and Computer Engineering, \\ Ingenuity Labs Research Institute}
	\streetaddress{}
	\city{Queen's University, Kingston}
	\state{}
	\country{Canada}
}
\email{ali.etemad@queensu.ca}
\email{}
\renewcommand{\shortauthors}{Behinaein, Bhatti, Rodenburg, Hungler, and Etemad}

\begin{abstract}
Electrocardiogram (ECG) has been widely used for emotion recognition. This paper presents a deep neural network based on convolutional layers and a transformer mechanism to detect stress using ECG signals. We perform leave-one-subject-out experiments on two publicly available datasets, WESAD and SWELL-KW, to evaluate our method. Our experiments show that the proposed model achieves strong results, comparable or better than the state-of-the-art models for ECG-based stress detection on these two datasets. Moreover, our method is end-to-end, does not require handcrafted features, and can learn robust representations with only a few convolutional blocks and the transformer component.
\end{abstract}



\maketitle

\section{Introduction}\let\thefootnote\relax\footnote{https://doi.org/10.1145/3460421.3480427}
Affective computing studies how machines can recognize, infer, process, and simulate human emotions \cite{picard2000affective}, with applications in education, health care, video games, and others \cite{picard2000affective, poria2017review, ross2019toward}. The ubiquitous availability of consumer-grade wearable sensing devices that collect biological signals (e.g., ECG) and the availability of deep learning frameworks have facilitated affective computing technologies \cite{koelstra2011deap, schmidt2018introducing, correa2018amigos}. 
Classical machine learning and feature engineering methods have been used to extract handcraft features and classify affect states \cite{ferdinando2016comparing, guo2016heart, schmidt2018introducing, bota2020emotion, bajpai2020evaluating, sriramprakash2017stress}.  Although handcrafted features perform well on emotion recognition, extracting them requires field expertise as such features are very application-specific. Convolutional layers have been employed to automate the feature extraction process \cite{hwang2018deep, sarkar2020self, sarkar2020self_icassp}. In addition, Transformer architectures \cite{vaswani2017attention} have recently emerged as a powerful solution and an alternative to recurrent neural networks for processing sequential data, and have been widely used in natural language processing \cite{devlin2018bert, wolf2020transformers} and computer vision \cite{khan2021transformers, han2020survey}. 

Due to the sequential nature of ECG time-series, transformers are viable candidates to learn spatio-temporal representations \cite{wang2021arrhythmia, che2021constrained}. In this paper, we propose an architecture that uses ECG to detect stress based on a combination of convolutional and transformer architectures. Our model uses only two convolutional blocks, which is considerably less compared to other works in the area \cite{sarkar2020self, lin2019explainable}. We test our proposed model on two publicly available affective computing datasets, WESAD \cite{schmidt2018introducing} and SWELL-KW \cite{koldijk2014swell}, using leave-one-subject-out (LOSO) scheme. Initial results using LOSO demonstrate that a fine-tuning (calibration) step is required to yield competitive results versus prior work. We demonstrate that by fine-tuning the model on only 10\% of user-specific data, strong results are achieved.

\section{Method}

\begin{figure}[t!]
	\centering
	\includegraphics[width=0.95\linewidth]{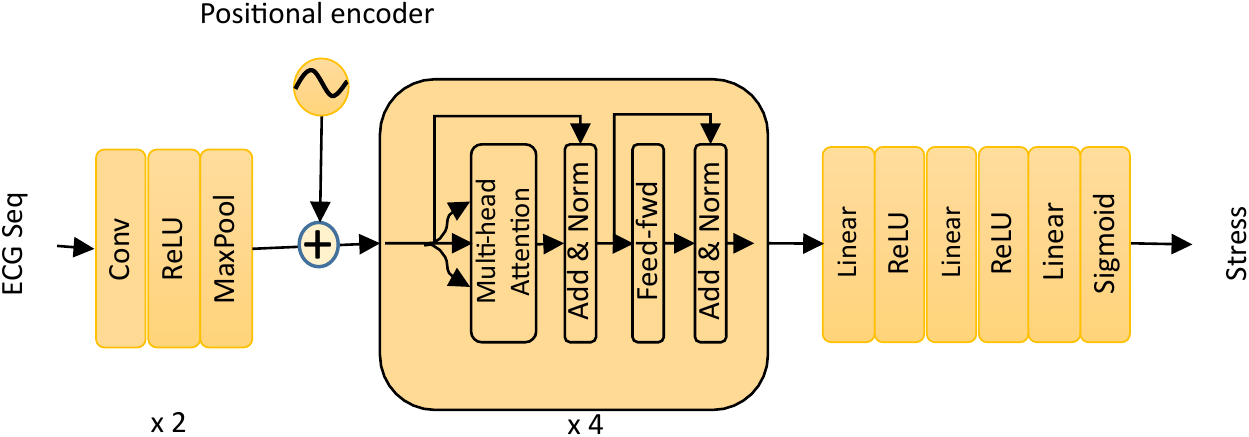}
	\caption{Transformer architecture for stress detection.} 
	\Description{Transformer architecture for affect classification.}
	 \label{fig:network}
\end{figure}

We propose an end-to-end network comprising three subnetworks, a convolutional subnetwork, a transformer encoder, and a fully connected (FC) subnetwork. The model and architectural details are depicted in Figure \ref{fig:network}. The convolutional front-end subnetwork comprises two convolutional layers, each directly followed by a ReLU activation and a maxpooling layer. The convolutional layers are followed by a reshape layer to flatten the last dimension. The role of the convolutional subnetwork is to extract spatio-temporal features from raw input ECG signals and feed them to the encoder. Since using a multi-head component (which will come later) results in loss of ordering in the input sequence, a piece of information needs to be added to the embeddings to give the encoder some sense of order. Here, we use the positional encoder proposed by Vaswani et al. \cite{vaswani2017attention} and add its output to the embeddings obtained from the reshape layer before supplying them to the transformer encoder. 
Next, the encoder consists of a multi-head, self-attention layer, followed by a dropout and a layer normalization, then a fully connected feed-forward network, and finally a dropout and a layer normalization. We use scaled dot-product attention \cite{vaswani2017attention} for our model where a query, key, and value vectors are generated. These vectors are created for each input by multiplying the input by $W_q$, $W_k$, and $W_v$, which are learned weight matrices for query, key, and value, respectively. The queries, keys, and values are individually stacked to create $Q$, $K$, and $V$, respectively. The attention values are then computed based on \cite{vaswani2017attention}, as
    $Attention(Q,K,V) = softmax(\frac{QK^T}{\sqrt{d_k}}) V,$
where $Q \in R^{n\times d_q}$, $K \in R^{n\times d_k}$, and $V \in R^{n\times v_k}$. Here, $n$ is the length of the sequence, and $d_q$, $d_k$, and $d_v$ are the embedding dimensions of $Q$, $K$, and $V$, respectively. The scaling factor $\sqrt{d_k}$ is added to mitigate the softmax's small gradient when its argument is very large. We use four of these components in the transformer encoder. Embeddings generated by the transformer encoder are flattened and fed to three FC layers where each of the first  two FC layers is followed by a Rectified Linear Unit (ReLU) activation function and the third one is followed by a sigmoid function for achieving binary classification. The network specifications are as follows: first convolution layer (filters = 64, kernel = 64, strides = 8), second convolution layer (filters = 128, kernel = 32, strides = 4), d$_{model}$ = 1024, K, V, Q’s dimension = 1024, d$_{ff}$ = 512, heads = 4, and finally the three FC layers' output dimensions = 512, 256, and 1, respectively. A dropout of 0.5 is used after each of the first two FC layers.

We implement the model in PyTorch on an NVIDIA TITAN RTX GPU. We use an Adam optimizer with an initial learning rate of 0.0001 and exponential decay of 0.985. We use the weighted binary cross-entropy loss to deal with the class imbalance. We train the model for 70 epochs with a batch size of 256. 
Due to significant inter-subject variability, we use a small portion of each test subject's data for calibrating (fining-tuning) the model. In this step, each model is trained for 40 epochs and fine-tuned for 30 epochs. 

\section{Experiments and Result}

\noindent \textbf{Datasets.} We use two publicly available datasets, Wearable Stress and Affect Detection (WESAD) \cite{schmidt2018introducing} and SWELL knowledge work (SWELL-KW) \cite{koldijk2014swell}, for evaluating our model. WESAD is a multimodal dataset collected from 15 subjects using wrist-worn and chest-worn wearable devices. The affect status of the subjects is also recorded in the dataset. We use the ECG data from the chest-worn device. In the SWELL-KW dataset, several modalities were recorded from 25 subjects. In this study, we only consider the ECG modality.

\noindent \textbf{Data Pre-processing.} We apply a 5th-order Butterworth high-pass filter with a cutoff frequency of 0.5 Hz on ECG similar to \cite{Makowski2021neurokit}. ECG was originally sampled at 700 Hz and 2048 Hz in WESAD and SWELL-KW, respectively. We down-sample signals from both datasets to 256 Hz for our study. ECG signals are normalized using user-specific z-score normalization \cite{sarkar2020self}. In terms of the output classes, WESAD recorded three affective states, neutral, stress, and amusement. For binary classification (stress vs. non-stress), we merge the neutral and amusement states into a `non-stress' state. For SWELL-KW, we use the `neutral' as the non-stress state and `time pressure' and `interruptions' as stress state.

\noindent \textbf{Validation Schemes.} To evaluate our model, we perform LOSO validation. We segment the data with a window size of 30 seconds and incremental steps of 1 second. In the fine-tuning step, we use 1\%, 5\%, and 10\% of data to calibrate the model. 

\noindent \textbf{Results.} 
The results for stress detection on WESAD is presented in Table \ref{tab:loso_results}. Our model obtains an accuracy of 80.4\% and F$_1$ score of 69.7\%, which are below the state-of-the-art results. By fine-tuning the model with only 10\% of the test data, the performance is considerably boosted to an accuracy of 91.1\%, outperforming other methods in Table \ref{tab:loso_results}.
\begin{table}
	\caption{Classification Results. TF: Transformer, SVM: Support Vector Machine, LDA: Linear Discriminant Analysis, QDA: Quadratic Discriminant Analysis, FT: Fine-Tuned.}
	\label{tab:loso_results}
	\small{
		\begin{tabular}{lllllll}
			\toprule
			Dataset & Ref & Method & Modality & Approach & Acc. & F$_1$\\
			\midrule\midrule
			WESAD & \cite{bota2020emotion} & QDA & ECG & LOSO & 85.7 & - \\
			& \cite{schmidt2018introducing} & LDA & ECG & LOSO & 85.4 & 81.3 \\
			& Ours	 & TF & ECG  & LOSO &80.4 & 69.7 \\
			& \textbf{Ours}	 & TF & ECG  & LOSO (FT) &\textbf{91.1} & \textbf{83.3} \\
			\midrule
			SWELL & \cite{koldijk2016detecting}& SVM & Multi & LOSO & 58.9 & -- \\
			& Ours	& TF & ECG & LOSO & 58.1 & 58.8 \\
			& \textbf{Ours}	& TF & ECG & LOSO (FT) & \textbf{71.6} & \textbf{74.2} \\
			\bottomrule
		\end{tabular}
	}
\end{table}
For the SWELL-KW dataset, as can be seen from Table \ref{tab:loso_results}, we achieve an accuracy of 58.1  and  F$_1$ score of 58.8 which are comparable to  \cite{koldijk2016detecting}. It should be noted, however, that \cite{koldijk2016detecting} uses multi-modal data (facial, posture, computer interactions, ECG, and EDA) as apposed to our uni-modal approach. Similar to WESAD, we observe that fine-tuning on only 10\% of data results in a considerable performance boost and an accuracy of 71.6\% (see Table \ref{tab:loso_results}). Table \ref{tab:loso_abelation} shows the performance when different percentages of test data are used for calibration. As can be seen, in WESAD, we need to use more than 1\% of the data in the fine-tuning step to considerably improve the result. However, for SWELL-KW, calibrating with even 1\% of data boosts the performance to outperform the baselines. While our approach yields promising results and is end-to-end (does not require hand-crafted features), the results indicate that to generalize to unseen subjects better than the state-of-the-art, it requires calibration with a small amount of data, which is considered a limitation of our work. Nonetheless, we believe our method demonstrates the potential for transformer architectures to be used in the area of affective computing.

\begin{table}[!t]
	\caption{Fine-tuning results. Values are in Acc (F$_1$) format.}
	\label{tab:loso_abelation}
		\begin{tabular}{lllll}
			\toprule
			Dataset & No Tuning & 1\% &  5\% &  10\% \\
			\midrule\midrule
			WESAD & 80.4 (69.7) & 81.6 (69.8) & 89.9 (80.8) & 91.1 (83.3)\\
			SWELL-KW & 58.1 (58.8) & 67.4 (69.7) & 68.3 (70.8) & 71.6 (74.2) \\
			\bottomrule
		\end{tabular}
\vspace{-0.4cm}
\end{table}

\section{Conclusion}

We presented a model based on convolutional and transformer architectures for detecting stress versus non-stress using ECG signals. To test our model, we used two publicly available datasets, WESAD and SWELL-KW. We showed that our model can achieve competitive results by using transformers with few convolutional layers. The results using LOSO validation showed that by fine-tuning the model with only a fraction of the test data (10\%), the proposed model can outperform the baseline methods.

\bibliographystyle{ACM-Reference-Format}
\bibliography{ref}

\end{document}